\documentclass[%
 reprint,
]{revtex4-2}

\usepackage{graphicx}
\usepackage{dcolumn}
\usepackage{bm}

\usepackage{amssymb}
\usepackage{mathtools}
\usepackage{graphicx}
\usepackage[colorlinks=true, allcolors=blue]{hyperref}
\usepackage{braket}
\usepackage{dsfont}
\usepackage{float}

\begin{document}

\preprint{APS/123-QED}

\title{Differences between quantum and classical adiabatic evolution}

\author{Cyrill Bösch}
\email{cb7454@cs.princeton.edu}
\affiliation{Institute of Geophysics, ETH Zurich, 8092 Zurich, Switzerland}
\affiliation{Department of Computer Science, Princeton University, Princeton, NJ 08540, USA }
\author{Andreas Fichtner}
\affiliation{Institute of Geophysics, ETH Zurich, 8092 Zurich, Switzerland}
\author{Marc Serra-Garcia}
\affiliation{AMOLF, Science Park 104, 1098 XG Amsterdam, The Netherlands}



\begin{abstract}

Adiabatic evolution is an emergent design principle for time modulated metamaterials, often inspired by insights from topological quantum computing such as braiding operations. However, the pursuit of classical adiabatic metamaterials is rooted in the assumption that classical and quantum adiabatic evolution are equivalent. We show that this is only true in the limit where the frequencies of all the bands are at infinite distance from $0$; and some instances of quantum adiabatic evolution, such as those containing zero modes, cannot be reproduced in classical systems. This is because mode coupling is fundamentally different in classical mechanics. We derive classical conditions to ensure adiabaticity and demonstrate that only under these conditions - which are different from quantum adiabatic conditions -, the single band Berry phase and Wilczek-Zee matrix for everywhere degenerate bands emerge as meaningful quantities encoding the geometry of classical adiabatic evolution. Finally, for general multiband systems we uncover a correction term in the non-Abelian gauge potential for classical systems.

\end{abstract}

\maketitle



\section{Introduction}

Quantum-classical analogies have grown into a vibrant research domain, particularly with the emergence of topological photonic \cite{lu2014topological} and phononic \cite{huber2016topological} metamaterials. More recently, classical metamaterials with time and space modulation \cite{nassar2018quantization, chen2019mechanical, galiffi2022photonics} achieve topological pumping \cite{grinberg2020robust,riva2020adiabatic, tanaka2023topological, rosa2019edge, kraus2012topological,zilberberg2018photonic, jurgensen2021quantized, wang2023smart, xia2021experimental, xu2020physical} and non-Abelian physics \cite{qiu2023minimal, jiang2021experimental, zhang2022non}, inspiring both fundamental research and novel information processing concepts. At the heart of these advancements lies quantum adiabatic evolution and the corresponding geometric quantities \cite{berry1984quantal, wilczek1984appearance,wilczek2012introduction, pancharatnam1956generalized,zanardi1999holonomic, budich2013adiabatic}. Until now, studies have assumed that classical adiabatic evolution is subject to gap conditions analogous to those in the quantum case, and the geometric quantities arising are equivalent  \cite{nassar2018quantization,barlas2020topological, riva2021adiabatic,riva2023adiabatic, santini2023elastic, karlivcic2023non,grinberg2020robust, xia2021experimental, xu2020physical, rosa2019edge,qiu2023minimal}. Our article challenges this assumption by demonstrating fundamental differences in the mode-coupling equations arising from the fact that a time-dependent Hermitian classical system is mapped to a time-dependent non-Hermitian quantum system. This has two consequences: (i) it introduces a correction term in the non-Abelian gauge potential and (ii) it implies an additional condition to ensure adiabaticity in classical systems, rendering concepts such as braiding of zero-modes untenable. Nevertheless, we identify a limit where these differences vanish, and classical adiabatic evolution approaches the quantum case.

\subsection{Quantum adiabatic evolution}
In the context of a slowly changing quantum Hamiltonian $H^q(t)$ with instantaneous eigenstates $\ket{n(t)}$ and instantaneous eigenvalues $\lambda_n(t)$, a quantum state initially prepared as $\ket{\psi^q(0)} = \ket{n(0)}$, and evolving under the Schrödinger equation $i \ket{\dot{\psi^q}} = H^q(t)\ket{\psi^q}$, remains in the $n$-th eigenstate, approximately described as $\ket{\psi_{adia}^q(t)} \approx \exp(-i\Lambda(t) - \gamma(t))\ket{n(t)}$ \cite{berry1984quantal}. Here, $\Lambda_n(t) := \int_0^t dt' \lambda_n(t')$ represents the dynamic phase, and $\gamma_n(t) = \int_0^t dt' \braket{n(t')|\dot{n}(t')}$ corresponds to the geometric phase, known as the Berry phase.  In what follows we choose a parallel transport gauge, i.e. $\braket{n(t)|\dot{n}(t)} = 0$. Then the Berry phase can be computed as $\gamma(t) = \operatorname{Im} \ln \braket{n(0)|n(t)}$  \cite{vanderbilt2018berry,samuel1988general}. To obtain $\ket{\psi_{adia}^q(t)}$, we expand $\ket{\psi^q(t)}$ in terms of the instantaneous eigenstates as $\ket{\psi^q(t)} = \sum_n q_m(t) \exp(-i\Lambda_m(t))\ket{m(t)}$. Substituting this into the Schrödinger equation and contracting with $\bra{n}$, we integrate from $0$ to $t$, leading to \cite{amin2009consistency,kato1950adiabatic, sakurai1995modern,born1928beweis}
\begin{equation}\label{eq: quantum coefficients}
\begin{split}
q_n(t) - q_n(0) = - \sum_{m \neq n} \int_0^t dt' q_m(t') \braket{n|\dot{m}} \\
\times e^{-i \int_0^{t'} dt''\lambda_{mn}(t'')},
\end{split}
\end{equation} 
where $q_n(0) = \delta_{nm}$ and $\lambda_{mn}(t) = \lambda_m(t)-\lambda_n(t)$. The right-hand side (RHS) can be interpreted as the error of the adiabatic approximation due to the excitation of other modes and is negligible if the exponential in the integrand oscillates rapidly compared to the slowly changing other terms. This condition is captured by the quantitative gap condition,
\begin{equation} \label{eq: quantum gap condition}
\max_{t \in [0,T_f]} \Bigg|\frac{\braket{n|\dot{m}}}{\lambda_{mn}(t)}\Bigg| \ll 1 \quad \forall n \neq m,
\end{equation}
where $T_f$ is the final time of integration.


\subsection{Classical adiabatic evolution}

In this letter we consider a classical evolution where a dynamical matrix $H^c(t)$ is changed slowly. $H^c(t)$ is symmetric, positive definite and real-valued as classical systems have real-valued couplings, but in principle the reasoning is also valid for complex-valued Hermitian matrices. The dynamics of the classical state vector $\ket{\psi^c(t)}$ is governed by the Newtonian second-order equation $\ket{\ddot{\psi^c}} = -H^c(t)\ket{\psi^c}$.
We denote the real-valued instantaneous frequency of $H^c(t)$ as $\omega_n(t) := \sqrt{\lambda_n(t)} > 0$, the dynamic phase as $\Omega_n(t) := \int_0^tdt' \omega_n(t')$ and we choose the instantaneous eigenstates real-valued and in a parallel transport gauge. As we will prove shortly, for an initial condition of $\ket{\psi^c(0)}= \ket{ n(0)}$ and $\ket{\dot{\psi}^c(0)}= 0$, the classical adiabatic approximation reads
\begin{equation} \label{eq: classical adiabatic evolution single band}
\ket{\psi_{adia}^c(t)} \approx \sqrt{ \frac{\omega_n(0)}{\omega_n(t)}} \cos\big(\Omega_n(t)\big)\ket{ n(t)}.
\end{equation}
We observe that the amplitude is not conserved and increases (decreases) as the mode becomes softer (stiffer). The Berry phase is hidden in the eigenstates and can be extracted as $\gamma(t) = \operatorname{Im} \ln \braket{n(0)|n(t)}$. For a closed-loop evolution, i.e. $H^c(T) = H^c(0)$, the Berry phase is obviously quantized to $0$ or $\pi$ modulo $2\pi$, protected by the fact that $H$ is purely real-valued and symmetric. 

\subsection{Problem statement and structure of the paper}

The expression \eqref{eq: classical adiabatic evolution single band} has also been obtained using WKB assymptotics in \cite{nassar2018quantization}. However, there and in other studies \cite{barlas2020topological, riva2021adiabatic,riva2023adiabatic, santini2023elastic, karlivcic2023non,grinberg2020robust, xia2021experimental, xu2020physical, rosa2019edge} it is assumed that the classical adiabatic condition is exclusively based on a frequency gap, analogous to the quantum gap condition \eqref{eq: quantum gap condition}. We will show, using a quantum formalism, that a gap condition \eqref{eq: quantum gap condition} is not sufficient and, more generally, quantum and classical adiabatic evolution are different except in the limit of infinitely high frequencies.

The study is based on contrasting the quantum coupling equations \eqref{eq: quantum coefficients}, with classical coupling equations, from which we deduce the adiabatic conditions and approximations. In Section \ref{s: derivation 1} we derive the coupling equations by applying a transformation that would symmetrize the problem in the time-independent case but fails to do so for time-dependent systems. In Section \ref{s: derivation 2} we explore the results in the light of biorthogonal operator theory. Then, from the coupling equations we derive classical adiabatic conditions and compare them to quantum conditions in Section \ref{s: classical vs quantum conditions}. We find fundamental differences and identify a limit where they become equivalent. In particular, setting $H^c \coloneqq H$ and $H^q \coloneqq \sqrt{H}$, we will show that, while generally different, in the limit where the full spectrum of $H$ is at infinite distance from $0$, both the non-adiabatic and adiabatic dynamics of $H^q$ and $H^c$ will be the same. In Section \ref{s: energy} we further study energy consumption of adiabatic evolution. We then present a simple numerical example in Section \ref{s: example}. Finally, in Section \ref{s: multiband} we generalize our findings to the multiband, non-Abelian case and discuss the appearance of a correction term to the non-Abelian gauge potential in classical systems in comparison to quantum systems. Conclusions are drawn in Section \ref{s: conclusion}.


\section{Derivation using coordinate transformation} \label{s: derivation 1}
It is well-known \cite{susstrunk2016classification, huber2016topological} that in time-independent systems, symmetric classical equations of motion can be mapped to a Hermitian Schrödinger equation. However, as we now demonstrate, this is not the case for time-dependent problems. The resulting Schrödinger equation is non-Hermitian because the coordinate transformation does not commute with the time derivative.

To see this we first define $\ket{\Psi^c(t)} = [\ket{\psi^c(t)},\ket{\dot{\psi}^c(t)}]^T$, such that the classical equation of motion can be transformed to first order
\begin{equation} \label{eq: rendered first order}
\ket{\dot{\Psi}^c} = \begin{bmatrix}
0 & \mathds{1} \\
-H(t) & 0
\end{bmatrix}\ket{\Psi^c}.
\end{equation}
We now introduce the transformation \cite{susstrunk2016classification, huber2016topological}
\begin{equation} \label{eq: transformation classical to quantum}
\mathcal{\mathcal{T}}(t) = \begin{bmatrix}
\sqrt{H(t)} & 0 \\
0 & i\mathds{1}
\end{bmatrix},
\end{equation}
which would symmetrize equation \eqref{eq: rendered first order}, i.e. mapping it to a Hermitian Schrödinger equation, for time-independent $H$.

Let us further define $\ket{\tilde{\Psi}^c} \coloneqq \mathcal{\mathcal{T}}(t)\ket{\Psi^c}$ and
\begin{equation}
\tilde{H}(t) := \begin{bmatrix}
0 & \sqrt{H(t)} \\
\sqrt{H(t)} & 0
\end{bmatrix}.
\end{equation}
Then, transforming equation \eqref{eq: rendered first order} under $\mathcal{T}(t)$, we obtain the equation of motion for $\ket{\tilde{\Psi}^c}$ as
\begin{equation} \label{eq: classical time depndent schrödinger}
i \ket{\dot{\tilde{\Psi}}^c} = \Big[\tilde{H}(t)+ i\dot{\mathcal{T}}(t) \mathcal{T}^{-1}(t)\Big]\ket{\tilde{\Psi}^c}.
\end{equation}
The inverse transformation, $\mathcal{T}^{-1}$, exists because $H$ is assumed to be positive definite.
This is a non-Hermitian Schrödinger equation because
\begin{equation}
i \dot{\mathcal{T}}(t) \mathcal{T}^{-1}(t) = i \begin{bmatrix}
\dot{\sqrt{H(t)}} \sqrt{H(t)}^{-1} & 0 \\
0 & 0
\end{bmatrix}
\end{equation}
is in general non-Hermitian. All the differences between quantum and classical adiabatic evolution and mode coupling arise from this term. We now expand the general (non-adiabatic) solution in the instantaneous, adiabatic eigenbases of $\tilde{H}$, i.e.
\begin{equation} \label{eq: expansion of classical state}
\ket{\tilde{\Psi}^c(t)} = \sum_{m, \sigma} c_{\sigma,m}(t) \sqrt{ \omega_m(0)\omega_m(t) } e^{-i \sigma \Omega_m(t)}\ket{\sigma, m(t)},
\end{equation}
where $\ket{\sigma, m(t)} := 1/\sqrt{2}[\ket{m(t)},\sigma \ket{m(t)}]^T$ are the orthonormal instantaneous eigenstates of $\tilde{H}(t)$ in a parallel transport gauge, i.e. $\braket{m(t),\sigma|\sigma, \dot{m}(t)} = \braket{m(t)| \dot{m}(t)} =  0$, and corresponding instantaneous eigenfrequencies are $\sigma \omega_m(t)$, with $\sigma = \pm 1$ labeling the subspace corresponding to the positive or negative eigenfrequencies \cite{susstrunk2016classification}. While for time-independent systems these two subspaces are always decoupled (as are eigenstates in general), in the time-dependent case this is no longer true. Inserting this expansion into equation \eqref{eq: classical time depndent schrödinger}, contracting with $\bra{m(t),\tau}$ and integrating from $0$ to $t$, one arrives at the classical version of equation \eqref{eq: quantum coefficients} (see Appendix \ref{ap: A})
\begin{widetext}
\begin{equation}\label{eq: classical mode coupling}
\begin{split}
c_{\tau,n}(t) - c_{\tau,n}(0) = - \sum_{m \neq n} \int_0^t dt' c_{\tau,m}(t') \braket{n|\dot{m}} \frac{1}{2} \sqrt{\frac{\omega_m(0)}{\omega_n(0)}} \Bigg(\sqrt{\frac{\omega_n(t')}{\omega_m(t')}} + \sqrt{\frac{\omega_m(t')}{\omega_n(t')}}\Bigg) e^{i \tau \Omega^-_{nm}(t')}\\
- \sum_{m \neq n} \int_0^t dt c_{-\tau,m}(t') \braket{n|\dot{m}} \frac{1}{2} \sqrt{\frac{\omega_m(0)}{\omega_n(0)}} \Bigg(\sqrt{\frac{\omega_n(t')}{\omega_m(t')}} - \sqrt{\frac{\omega_m(t')}{\omega_n(t')}}\Bigg) e^{-i (-\tau) \Omega^+_{nm}(t')}\\
+ \int_0^t dt c_{-\tau,n}(t') \frac{1}{2} \frac{\dot{\omega}_n(t')}{\omega_n(t')}e^{i\tau 2 \Omega_n(t')}, 
\end{split}
\end{equation}
\end{widetext}
where $c_{\tau,n}(0) = \delta_{nm}\delta_{\tau\sigma}\braket{n(0),\tau|\mathcal{T}(0)|\Psi^c(0)}/\omega_n(0)$,  $\Omega^-_{nm}(t) := \Omega_n(t)-\Omega_m(t)$ and $\Omega^+_{nm}(t) := \Omega_n(t)+\Omega_m(t)$. Again, the RHS can be identified as the error to the adiabatic approximation. Clearly, if the RHS is approximately 0, so is the left-hand side, i.e. $c_{\tau,n}(t) \approx c_{\tau,n}(0)$. Substituting this result back into \eqref{eq: expansion of classical state}, applying the inverse transformation, i.e. $\mathcal{T}^{-1}(t)\ket{\tilde{\Psi}^c(t)}$ and taking the real part, results in the adiabatic approximation \eqref{eq: classical adiabatic evolution single band}. 

Because the transformation does not commute with the time derivative operator, Hermitian classical systems fundamentally map to non-Hermitian quantum systems giving rise to differences in the (adiabatic) evolution for finite frequencies, as analyzed in Section \ref{s: classical vs quantum conditions}.

\section{Derivation using biorthogonal formalism} \label{s: derivation 2}
It is customary to treat non-Hermitian systems in the biorthogonal formalism, directly starting from equation \eqref{eq: rendered first order} \cite{ibanez2014adiabaticity}. Now we show that such a treatment produces the same coupling equations \eqref{eq: classical mode coupling} which we demonstrate to be well-defined despite the time-dependent norm ambiguity of biorthogonal basis functions.

Let us define
\begin{equation}\label{ap2 eq: non Herm op}
   G :=  \begin{bmatrix}
0 & \mathds{1} \\
-H(t) & 0
\end{bmatrix},
\end{equation}
such that equation \eqref{eq: rendered first order} reads
\begin{equation} \label{eq: rendered first order}
\ket{\dot{\Psi}^c} = G\ket{\Psi^c}.
\end{equation}
The $m$-th instantaneous eigenvalue-eigenstate pair of $H(t)$ is denoted by $(\lambda_m(t) = \omega^2_m(t), \ket{m(t)})$. Without loss of generality, we again choose $\ket{m(t)}$ in a parallel transport gauge such that $\braket{m|\dot{m}} = 0$. Let us introduce the biorthogonal left and right eigenvectors of $G$ \cite{muga2004complex}. It is easy to check that
\begin{equation}\label{ap2 eq: right eigvecs}
\begin{split}
   \ket{\sigma, M(t)} := \frac{1}{\sqrt{2}} \begin{bmatrix}
        \ket{m(t)} \\
        -\sigma i\omega_m(t) \ket{m(t)}
    \end{bmatrix}, \\
    \quad \beta_m(t) := -\sigma i\omega_m(t)
\end{split}
\end{equation}
for $\sigma = \pm 1$ is a right eigenstate-eigenvalue pair of the non-Hermitian operator \eqref{ap2 eq: non Herm op}. Its corresponding left eigenstate is given by 
\begin{equation}\label{ap2 eq: left eigvecs}
\bra{ \hat{N}(t), \tau} :=\frac{1}{\sqrt{2}} 
\begin{bmatrix}
    \bra{n(t)} &
     \frac{i\tau}{\omega_n(t)} \bra{n(t)}
\end{bmatrix},
\end{equation}
with corresponding eigenvalue
\begin{equation}
    \hat{\beta}_n(t) = \beta^*_n(t) := \tau i\omega_n(t),
\end{equation}
where $*$ denotes complex conjugation. Note that because $H(t)$ is assumed to be positive definite and symmetric, we have $\lambda_m(t) >  0$ and hence $\omega_m(t) \subset \mathbb{R}^{+}$, implying that the instantaneous non-Hermitian system is purely oscillatory. 

We further note that degeneracies in $H$ carry over to degeneracies in $G$. We assume that respective eigenstates of $H$ have been orthogonalized so that corresponding left and right eigenstates are naturally biorthogonal.
The left and right eigenstates satisfy the biorthogonality relation 
\begin{multline}\label{ap2 eq: biorthogonality relation}
    \braket{\hat{N}(t), \tau |\sigma, M(t)} \\
    = \frac{1}{2}\big(\braket{n|m}+\tau \sigma \frac{\omega_m(t)}{\omega_n(t)}\braket{n|m}\big)
    = \delta_{nm} \delta_{\sigma \tau}.
\end{multline}

Biorthogonal eigenstates of non-Hermitian operators do not only have an arbitrary phase but also an arbitrary norm \cite{leclerc2012role,ibanez2014adiabaticity}. Hence, for any at least once differentiable $f_{\sigma,m}(t) \in \mathbb{C}$ the right $f$-transformed eigenstates $\ket{\sigma, M(t)} \rightarrow f_{\sigma,m}(t)\ket{\tau, M(t)}$ and left $f$-transformed eigenstates $\bra{\hat{N(t)},\tau} \rightarrow \bra{\hat{N}(t),\tau}1/f_{\tau,n}^*(t)$ also form a complete biorthogonal basis. Furthermore, because here we are interested in parallel transporting $\ket{m}$ rather than $\ket{\tau, M(t)}$ in general there is no further restriction on $f_{\sigma,m}(t)$ (see \cite{ibanez2014adiabaticity} for more in detail discussion). 


To that end we use the following non-adiabatic ansatz
\begin{multline}
\label{eq: expansion of classical state 2}
\ket{\Psi^c(t)} = \sum_{m, \sigma} b_{\sigma,m}(t)e^{-i \sigma \Omega_m(t)} f_{\sigma,m}(t) \ket{\sigma, M(t)}.
\end{multline}
We substitute the ansatz into the equations of motion \eqref{eq: rendered first order} and contract with $\bra{\hat{N},\tau}1/f^*_{\tau,n}$, where from now on we omit to show the time dependence explicitly. We obtain (see Appendix \ref{ap: B})
\begin{multline}\label{eq: coupling equations for b}
    \dot{b}_{\tau,n} f_{\tau,n}\sqrt{\omega_n} = -b_{\tau,n}\dot{f}_{\tau,n}\sqrt{\omega_n} 
    -b_{\tau,n}f_{\tau,n}\frac{1}{2}\frac{\dot{\omega}_n}{\sqrt{\omega_n}} \\
    - \sum_m b_{\tau,m}f_{\tau,m}\braket{n|\dot{m}}\frac{1}{2}\sqrt{\omega_n}\Big(1+ \frac{\omega_m}{\omega_n}\Big) e^{i\tau\Omega^-_{nm}}\\
    - \sum_m b_{-\tau,m}f_{-\tau,m}\braket{n|\dot{m}}\frac{1}{2}\sqrt{\omega_n}\Big(1- \frac{\omega_m}{\omega_n}\Big) e^{i\tau\Omega^+_{nm}}\\
    +b_{-\tau,n}f_{-\tau,n}\frac{1}{2}\frac{\dot{\omega}_n}{\sqrt{\omega_n}}e^{i2\tau\Omega_n}.
\end{multline}

To demonstrate that the results obtained in this study are well-defined we have to show that the coupling equations \eqref{eq: coupling equations for b} are independent of the arbitrary choice of $\{f_{\sigma,m}\}$. We will show that this is in indeed the case, as the choice of $\{f_{\sigma,m}\}$ will just turn out to be a time-dependent scaling which is compensated by the corresponding coefficients.

To that end we define $a_{\tau,n} := b_{\tau,n} f_{\tau,n} \sqrt{\omega_n}$ with
\begin{equation}
    \dot{a}_{\tau,n} = \dot{b}_{\tau,n} f_{\tau,n} \sqrt{\omega_n} + b_{\tau,n}f_{\tau,n} \frac{1}{2}\frac{\dot{\omega_n}}{\sqrt{\omega}_n} + b_{\tau,n} \dot{f}_{\tau,n}\sqrt{\omega_n}.
\end{equation}
We further use $b_{\tau,n} = a_{\tau,n}/(f^*_{\tau,n}\sqrt{\omega_n})$ and because
\begin{equation}
    \dot{b}_{\tau,n} f_{\tau,n}  \sqrt{\omega_n} = \dot{a}_{\tau,n} - a_{\tau,n} \frac{1}{2}\frac{\dot{\omega}_n}{\omega_n} - a_{\tau,n} \frac{\dot{f}_{\tau,n}}{f^*_{\tau,n}},
\end{equation}
all the diagonal terms in equation \eqref{eq: coupling equations for b} vanish and all $f$-terms are absorbed into the coefficients $a_{\tau,n}$ such that the coupling equations for $a_{\tau,n}$ read
\begin{multline}\label{ap2 eq: coupling equations for a}
    \dot{a}_{\tau,n}  = 
    - \sum_m a_{\tau,m}\braket{n|\dot{m}}\frac{1}{2}\Big(\sqrt{\frac{\omega_n}{\omega_m}}+ \sqrt{\frac{\omega_m}{\omega_n}}\Big) e^{i\tau\Omega^-_{nm}}\\
    - \sum_m a_{-\tau,m}\braket{n|\dot{m}}\frac{1}{2}\Big(\sqrt{\frac{\omega_n}{\omega_m}}- \sqrt{\frac{\omega_m}{\omega_n}}\Big) e^{i\tau\Omega^+_{nm}} \\
     +a_{-\tau,n}\frac{1}{2}\frac{\dot{\omega}_n}{\omega_n}e^{i2\tau\Omega_n}.
\end{multline}

We note that the above equations are equivalent to \eqref{eq: classical mode coupling} up to factors of $\sqrt{\omega_n(0)/ \omega_m(0)}$ which can be pulled out and absorbed in the coefficients.

Therefore, upon time integration, these coupling equations for $\{a_{\tau,n}(t)\}$ are equivalent to equation \eqref{eq: classical mode coupling} for the coefficients $\{c_{\tau,n}(t)\}$. The ambiguous scaling factors $\{f_{\sigma,m}\}$ do not appear in the coupling equation, meaning that they are well-defined.

The choice of $\{f_{\sigma,m}\}$ results in a scaling of the coefficients that leave the overall, generally non-adaibatic state invariant. To see this, observe that $\mathcal{T}(t) \sqrt{\omega_m(0)/\omega_m(t)}\ket{\sigma, M(t)} = \sqrt{\omega_m(0)\omega_m(t)}\ket{\sigma, m(t)}$. Therefore in Section \ref{s: derivation 1} we use the biorthogonal basis with $\hat{f}_{\tau,n}(t) = \hat{f}_n(t) = \sqrt{\omega_m(0)/\omega_m(t)}$ and transform it under $\mathcal{T}(t)$. Hence, if we back-transform the ansatz \eqref{eq: expansion of classical state} with $\mathcal{T}^{-1}(t)$ we get 
\begin{equation}
\ket{\Psi^c(t)}
= \sum_{m, \sigma} c_{\sigma,m}(t) \sqrt{ \frac{\omega_m(0)}{\omega_m(t)}} e^{-i \sigma \Omega_m(t)}\ket{\sigma, M(t)}.
\end{equation}
Comparing this ansatz with the one chosen in \eqref{eq: expansion of classical state 2}, we find that the different coefficiencts are related by a scaling of $b_{\sigma,m}(t)f_{\sigma,m}(t)\sqrt{\omega_m(t)/\omega_m(0)} = a_{\sigma,m}(t)/\sqrt{\omega_m(0)} = c_{\sigma,m}(t)$. Solving for $b_{\sigma,m}$ and substituting the expression back into \eqref{eq: expansion of classical state 2}, we see that $f_{\sigma,m}$ vanishes and therefore conclude that $\ket{\Psi^c(t)}$ is independent of the choice of $\{f_{\sigma,m}\}$, as expected.

In summary, a choice of $\{f_{\sigma,m}\}$, here $\{\hat{f}_{\sigma,m}\}$, results in a unique set of coefficients. The coupling between these coefficients is independent of $\{f_{\sigma,m}\}$. Furthermore, while the time-dependent scaling of the coefficients depends on $\{f_{\sigma,m}\}$, $\ket{\Psi^c(t)}$ does not. In Section \ref{s: derivation 1}, the coefficients $\{c_{\sigma,m}\}$ were chosen such that in the adiabatic limit they remain constant. 

An extensive discussion about the scaling issue, which is about how to define the population of biorthogonal eigenstates, can be found in \cite{ibanez2014adiabaticity}. In this article we focus on the mode coupling, which is well-defined, despite the ambiguity of the norm of the biorthogonal eigenstates.

\section{Classical vs. quantum adiabatic evolution: differences and a limit where they meet} \label{s: classical vs quantum conditions}

\begin{figure*}
\includegraphics[width=1\linewidth]{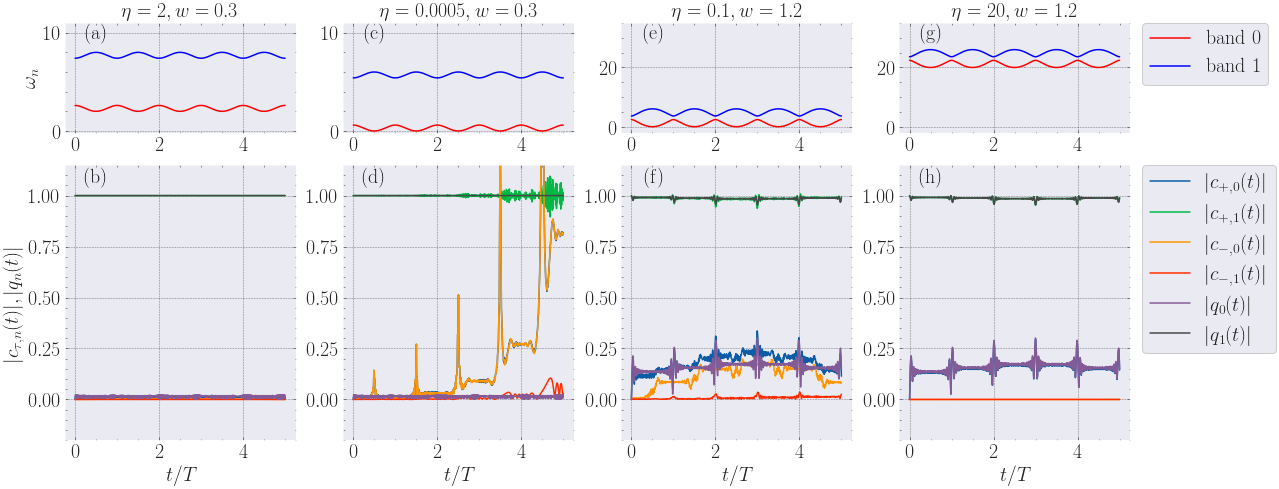}
\caption{\label{fig: bands and couplings}Numerical simulation of a $2\times2$ system defined in equation \eqref{eq: example hamiltonain}, where $H^c = h^2$ and $H^q = h$. Row 1 displays the band structures, while row 2 shows the time evolution of the classical and quantum mode coefficients. The coefficients are initialized in the excited mode for various values of $\eta$ and $w$, while $s=3$ and $T=10$ are fixed. (a, b) In the presence of a large gap and the absence of a zero mode, both the classical and quantum systems exhibit adiabatic behavior. (c, d) With a small amount of energy transitioning to the lower mode, the classical system violates adiabatic conditions due to interactions between $c_{+,0}$ and $c_{-,0}$. In contrast, the quantum system remains adiabatic. (e-h) The classical system approaches quantum behavior as the spectrum is shifted to higher frequencies. In (e, f), non-adiabatic interactions occur due to the small band gap (quantum) and the combination of a small gap and a near-zero mode (classical). Shifting the spectrum to higher frequencies in (g, h) results in the classical system where only the gap matters, and the frequency ratios approach unity. Consequently, the modes from the negative frequency axis are no longer excited, and the full non-adiabatic dynamics closely approximates that of the quantum system.}
\end{figure*}

We now analyse the differences between quantum and classical adiabatic evolution by comparing the RHS of the classical coupling equations \eqref{eq: classical mode coupling} and \eqref{ap2 eq: coupling equations for a} to the RHS of the quantum mechanical couplings \eqref{eq: quantum coefficients}: (i) The first RHS-term in equation \eqref{eq: classical mode coupling} resembles the quantum mechanical coupling to other modes of the same subspace $\tau$. In the classical case, however, the square root ratio of the tracked and all the other instantaneous frequencies scales this error term. If any mode (tracked or not tracked) approaches a zero frequency, this term diverges, requiring ever slower time evolution of $H(t)$. (ii) The second sum is the error due to coupling to other modes of the other subspace, denoted with $-\tau$. It will be clear shortly that if the first term is negligibly small, so is the second term. (iii) Finally, the last term is the coupling to the same mode of the other subspace which can lead to amplitude pumping through parametric amplification. Because this last term holds simultaneously for all $c_{\tau,n}$, small transition errors to a near-zero mode, $c_{\tau,0}$, may grow over time. This is exactly the situation in Figure \ref{fig: bands and couplings} (c,d), as discussed shortly. 

Following the same logic used to derive the quantum gap condition \eqref{eq: quantum gap condition}, we can now establish two classical conditions ensuring that the integrals in equation \eqref{eq: classical mode coupling} become negligible thus implying the validity of the adiabatic approximation \eqref{eq: classical adiabatic evolution single band}:
\begin{equation}\label{eq: classical adiabatic condition 1}
\begin{split}
\max_{t \in [0,T_f]} \Bigg|\frac{\braket{n|\dot{m}}}{\omega^-_{mn}(t)} \frac{1}{2} \sqrt{\frac{\omega_m(0)}{\omega_n(0)}} \Bigg(\sqrt{\frac{\omega_n(t)}{\omega_m(t)}} + \sqrt{\frac{\omega_m(t)}{\omega_n(t)}}\Bigg)\Bigg|  \ll 1 \\ \forall m \neq n,
\end{split}
\end{equation}
and
\begin{equation}\label{eq: classical adiabatic condition 2}
\max_{t \in [0,T_f]} \Bigg|\frac{1}{4} \frac{\dot{\omega}_n(t)}{\omega^2_n(t)}\Bigg| \ll 1.
\end{equation}
We stress that condition \eqref{eq: classical adiabatic condition 1} implies also that the second sum in equation \eqref{eq: classical mode coupling} is small because the absolute difference of the positive ratios is smaller than their sum and the same applies to the more rapid oscillations, i.e., $\omega^-_{nm}(t) = \omega_n(t)-\omega_m(t) < \omega^+_{nm}(t) = \omega_n(t)+\omega_m(t)$. The first condition resembles the quantum mechanical gap condition, however scaled by the frequency ratios. The second condition is a condition on the distance from the zero frequency line.

Note that for a single degree of freedom (DOF) harmonic oscillator there is only one coupling to its mirrored mode partner, the last line in equation \eqref{eq: classical mode coupling}. In Appendix \ref{ap: C} we apply our study to the single DOF case in detail.

In equations \eqref{eq: classical mode coupling} we can identify the limit $\omega_n \rightarrow \infty, \, \forall n$ where the classical time-dependent dynamics under $H(t)$ approach the dynamics of a quantum system under $\sqrt{H(t)}$. In this limit $\sqrt{\omega_n(t)/\omega_m(t)} \rightarrow 1$ which implies that
\begin{equation}
    \frac{1}{2} \sqrt{\frac{\omega_m(0)}{\omega_n(0)}} \Bigg(\sqrt{\frac{\omega_n(t')}{\omega_m(t')}} + \sqrt{\frac{\omega_m(t')}{\omega_n(t')}}\Bigg) \rightarrow 1,
\end{equation}
and 
\begin{equation}
    \frac{1}{2} \sqrt{\frac{\omega_m(0)}{\omega_n(0)}} \Bigg(\sqrt{\frac{\omega_n(t')}{\omega_m(t')}} - \sqrt{\frac{\omega_m(t')}{\omega_n(t')}}\Bigg) \rightarrow 0.
\end{equation}
Finally, the last line in equation \eqref{eq: classical mode coupling} approaches $0$, therefore the coupling dynamics reduces to the quantum case \eqref{eq: quantum coefficients} and the classical adiabatic conditions to the quantum conditions. Note that this limit requires all the bands to be at sufficiently large frequencies, it does not apply to a subset of bands at high frequencies. In general, any arbitrary high frequency mode can couple to any near zero mode.


\section{Energy consumption} \label{s: energy}
The energy consumption of adiabatic driving is nontrivial as the norm is not conserved. From the classical adiabatic approximation \eqref{eq: classical adiabatic evolution single band} we see that the average instantaneous energy for a system with time-independent, unit mass is given by $\braket{E(t)} = \frac{1}{4}\omega_n(0)\omega_n(t)$. Therefore, the adiabatic energy cost of transforming an initial state $\ket{n(0)}$ into a final state $\ket{n(T)}$ is $\Delta E = \frac{1}{4}\omega_n(0)(\omega_n(T)-\omega_n(0))$, with $\Delta E = 0$ for a flat band. Furthermore, for a closed loop where $\omega_n(T) = \omega_n(0)$, the total energy cost is also zero, meaning that adiabatically accumulating a Berry phase does not consume any energy.

\section{Example}\label{s: example}

The following example illustrates the phenomena discussed above using a simple $2\times2$ system. To this end, we slightly reformulate the system considered in \cite{kariyado2016hannay}. We define the following matrix
\begin{widetext}
\begin{equation} \label{eq: example hamiltonain}
h(t) := \begin{bmatrix}
\eta + s + [w+(s-w)\cos(t/T+\pi)] & (s-w)\sin(t/T+\pi) \\
(s-w)\sin(t/T+\pi) & \eta + s - [w+(s-w)\cos(t/T+\pi)]
\end{bmatrix},
\end{equation}
\end{widetext}
and identify the quantum and classical systems as $H^q = h$ and $H^c = h^2$, respectively (hence, $H^q = \sqrt{H^c}$). By squaring the operator, the eigenstates and hence the Berry phase remain unchanged, while the eigenvalues are squared. This implies that the quantum and classical states oscillate with the same frequency. The parametrization of $h$ is chosen such that changing $\eta$ allows us to shift the spectrum without altering the gap. Furthermore, for $0<w<s/2$ $(s>w>s/2)$, the first (second) band has a Berry phase of $\pi$ $(0)$. The transition from $0$ to $\pi$ of both bands occurs at $w = s/2$, where the gap closes. Finally, $T$ controls the slowness of integration. The parameters $s$ and $T$ are fixed to $3$ and $10$, respectively. The classical system has four modes $\ket{+,0}, \ket{+,1}, \ket{-,0},\ket{-,1}$ with corresponding coefficients $c_{+,0}, c_{+,1}, c_{-,0}$ and $c_{-,1}$. The quantum modes are $\ket{0}$ and $\ket{1}$ with coefficients $q_0$ and $q_1$. In the following, we always initialize the system in the excited mode, $\ket{+,1}$ and $\ket{1}$, respectively, setting $c_{+,1}(0) = q_1(0) = 1$ and $c_{+,0}(0)=c_{+,1}(0)=c_{-,0}(0)=c_{-,1}(0)=q_0(0)=0$. We numerically integrate equations \eqref{eq: quantum coefficients} and \eqref{eq: classical mode coupling} and display the norm of the coefficients over $5$ periods in Figure \ref{fig: bands and couplings}. In panels (a,b), we set $\eta = 2, w=0.3$. We observe that the adiabatic approximation is well justified for both the quantum (due to the large gap) and the classical (due to the large gap and no near-zero mode) system. In (c,d), we set $\eta = 0.0005$, thereby shifting the lower band closer to zero. This has, as expected, no effect in the quantum case, which remains adiabatic. However, in the classical case, the small amount of energy that scatters to $c_{+,0}(t)$ and $c_{-,0}(t)$ amplifies because the second condition \eqref{eq: classical adiabatic condition 2} is violated for $\ket{+,0}$ and $\ket{-,0}$. Finally, (e-h) demonstrate how the classical system approaches the quantum system by shifting the spectrum to higher frequencies: In (e,f), the parameters are chosen as $\eta = 0.1,w=1.2$. The larger $w$ and hence smaller gap result in a non-adiabatic transition to the lower mode. Because the lower band is still close to zero, the classical coupling to the other modes, particularly the non-vanishing coupling to the negative frequency subspace, presents a different picture in the classical than in the quantum case. As predicted, shifting the spectrum to large enough frequencies (g-h) such that $\sqrt{\omega_0(t)/\omega_1(t)} \rightarrow 1$, the classical mode coupling behavior approaches that of the quantum system, which is, once again, left invariant by a spectrum shift.

\section{Multiband case} \label{s: multiband}
We now treat the case of generally interacting bands and also reduce it to the case where a set of bands are everywhere degenerate. Let us first recall the quantum behavior: Let $\mathcal{S}$ be a subspace of interacting bands that are well separated by an energy gap from all other bands such that 
\begin{equation} \label{eq: multiband quantum gap condition}
         \max_{t \in [0,T_f]} \Bigg|\frac{\braket{s|\dot{m}}}{\lambda_{sm}(t)}\Bigg| \ll 1 \quad \forall s \in \mathcal{S}, \forall m  \not\in\mathcal{S},
\end{equation}
then equation \eqref{eq: quantum coefficients} can be solved for the coefficients of the subspace, $\mathbf{q}_{\mathcal{S}}$, using the evolution operator $\mathcal{U}^q(t) \coloneqq \mathcal{U}^q(t,0)$ \cite{wilczek2012introduction, wilczek1984appearance}, i.e.
    \begin{equation}
        \mathbf{q}_{\mathcal{S}}(t) = \mathcal{U}^q(t)\mathbf{q}_{\mathcal{S}}(0),
    \end{equation}
with
\begin{equation}
        \mathcal{U}^q(t) = \mathcal{P} \exp \Big(-\int_0^tdt' \mathcal{Q}(t')A(t')\mathcal{Q}^{-1}(t')\Big),
\end{equation}
where $\mathcal{P}$ denotes the path ordering operator and $A_{ss'} := \braket{s|\dot{s}'}$ denotes the non-Abelian gauge field which is transformed under $\mathcal{Q}_{ss'}(t) := \exp(i\Lambda_s(t))\delta_{ss'}$.  In the case of everywhere degenerate bands one obtains $\mathcal{Q}(t)A(t)\mathcal{Q}^{-1}(t) \rightarrow A(t)$ and $\mathcal{U}^q(t) \rightarrow \mathcal{U}(t) = \mathcal{P} \exp(-\int_0^tdt' A(t'))$ is the Wilczek-Zee matrix whose trace is the gauge invariant Wilson loop \cite{wilson1974confinement}. To treat the classical system we denote coefficients of the particular frequency subspace, $\tau$, corresponding to states of the subspace as $[\mathbf{c}^{\tau}_{\mathcal{S}}]_s = c_{\tau,s}$. We introduce the classical evolution operator for the coefficients, $\mathcal{U}^c(t)\coloneqq \mathcal{U}^c(t,0)$, such that
\begin{equation}\label{eq: time evolution of classical subspace coefficients}
        \mathbf{c}^{\tau}_{\mathcal{S}}(t) = \mathcal{U}^c(t)\mathbf{c}^{\tau}_{\mathcal{S}}(0).
\end{equation}
In Appendix \ref{ap: A} we show that
\begin{equation}\label{eq: classical time evolution operator}
        \mathcal{U}^c(t) = \mathcal{P} \exp \Big(-\int_0^tdt' \mathcal{C}(t')\Big[A(t')+A^{\Delta}(t')\Big]\mathcal{C}^{-1}(t')\Big),
\end{equation}
with the classical transformation $\mathcal{C}_{ss'}(t) := \exp(i\Omega_s(t))\delta_{ss'}/\sqrt{\omega_s(0)\omega_s(t)}$ and $A^\Delta_{ss'}(t) := \braket{s|\dot{s}'}(\omega_s-\omega_{s'})/(2\omega_{s'})$ which can be seen as a classical correction to the non-Abelian gauge field and again reflects the fact that the coupling between interacting modes depends on their frequencies. As shown in Section \ref{s: derivation 2} the couplings and hence the classical non-Abelian gauge potential $A(t')+A^{\Delta}(t')$ is independent of the norm ambiguity of the biorthogonal basis. This implies that it is well-defined.

The high frequency limit can be identified as: $(\omega_s-\omega_{s'})/(2\omega_{s'}) \rightarrow 0$ implying that $A^\Delta_{ss'}(t) \rightarrow 0$ and $\sqrt{\omega_s(0)\omega_s(t)}/\sqrt{\omega_{s'}(0)\omega_{s'}(t)} \rightarrow 1$, implying that $\mathcal{C}(t)A(t)\mathcal{C}^{-1}(t)\rightarrow \mathcal{Q}(t)A(t)\mathcal{Q}^{-1}(t)$ hence the classical evolution approaches the evolution of the quantum system governed by $\sqrt{H}(t)$.  Substituting \eqref{eq: time evolution of classical subspace coefficients} back into the expansion \eqref{eq: expansion of classical state} and applying the inverse transformation $\mathcal{T}^{-1}(t)$ yields the classical, general multiband solution
\begin{multline} 
\label{eq: classical adiabatic evolution multiband band, general}
\psi_{adia,mult}^c(t) \approx \\
\operatorname{Re} \left\{ \sum_{s \in \mathcal{S}} \sum_{s' \in \mathcal{S}}\mathcal{U}^c_{ss'}(t)  c_{s'}(0)\sqrt{\frac{\omega_s(0)}{\omega_s(t)}} e^{-i \tau \Omega_s(t)} \ket{ s(t)}\right\},
\end{multline}
with initial coefficients  $c_{s'}(0)= \braket{s'(0),\tau|\mathcal{T}(0)|\Psi^c(0)}/\omega_{s'}(0) $. 
It remains to state the adiabatic conditions under which the approximation is valid. The first two are a restatement of the single band case but for the whole suspace $\mathcal{S}$,
\begin{equation} \label{eq: adiabatic condition multiband 1}
\begin{split}
\max_{t \in [0,T_f]} \Bigg|\frac{\braket{s|\dot{m}}}{\omega^-_{ms}(t)} \frac{1}{2} \sqrt{\frac{\omega_m(0)}{\omega_s(0)}} \Bigg(\sqrt{\frac{\omega_s(t)}{\omega_m(t)}} + \sqrt{\frac{\omega_m(t)}{\omega_s(t)}}\Bigg)\Bigg| \ll 1 \\ 
 \quad \forall s \in \mathcal{S}, \forall m  \not\in\mathcal{S},
\end{split}
\end{equation}
and
\begin{equation}\label{eq: adiabatic condition multiband 2}
\max_{t \in [0,T_f]} \Bigg|\frac{1}{4} \frac{\dot{\omega}_s(t)}{\omega^2_s(t)}\Bigg| \ll 1 \quad \forall s \in \mathcal{S}.
\end{equation}
Finally, we have to explicitly require that the interaction of bands in the subspace is isolated from the same subspace on the negative frequency axis, $-\tau$, which is ensured if
\begin{equation}\label{eq: adiabatic condition multiband 3}
\begin{split}
\max_{t \in [0,T_f]} \Bigg|\frac{\braket{s|\dot{s}'}}{\omega^+_{s's}(t)} \frac{1}{2} \sqrt{\frac{\omega_{s'}(0)}{\omega_{s}(0)}} \Bigg(\sqrt{\frac{\omega_s(t)}{\omega_{s'}(t)}} - \sqrt{\frac{\omega_{s'}(t)}{\omega_s(t)}}\Bigg)\Bigg| \ll 1 \\ \quad \forall s, s' \in \mathcal{S}.
\end{split}
\end{equation} 

In the case of everywhere degenerate bands we have $A^\Delta_{ss'}(t) = 0$ since $\omega_s = \omega_{s'}$ and $\mathcal{C}(t)A(t)\mathcal{C}^{-1}(t) \rightarrow A(t)$ which is integrated to yield, as in the quantum system, the Wilczek-Zee matrix, $\mathcal{U}(t)$. Defining $\omega_{\mathcal{S}}$ as the degenerate eigenfrequency of the subspace and $\Omega_\mathcal{S}$ the corresponding dynamical phase and again using that the eigenstates can always be chosen real for a real-valued symmetric system, one obtains
\begin{multline} 
\label{eq: classical adiabatic evolution multiband band, degenerate}
\psi_{adia,mult,deg}^c(t) \approx 
\\
\frac{\sqrt{\omega_\mathcal{S}(0)}}{\sqrt{\omega_\mathcal{S}(t)}} \cos\Big( \Omega_\mathcal{S}(t)\Big) \sum_{s \in \mathcal{S}} \sum_{s' \in \mathcal{S}}\mathcal{U}_{ss'}(t) c_{s'}(0)_s\ket{ s(t)},
\end{multline}
and the third condition becomes trivially satisfied.

\section{Conclusions} \label{s: conclusion}
We have demonstrated that a time-dependent Hermitian classical system is equivalent to a time-dependent non-Hermitian quantum system, resulting in distinct mode coupling equations. Using non-Hermitian operator theory we showed that these coupling equations are well-defined.

In contrast to the time-independent case, classical time-dependent dynamics can in general not be symmetrized because the transformation does not commute with the time derivative. From this we deduced that the quantum gap condition is not sufficient to ensure adiabaticity and certain quantum adiabatic phenomena can not be realized in classical systems. Another consequence of this is the appearance of a classical correction term to the non-Abelian gauge field for general multiband systems which deserves further investigation and may inspire a new class of time modulated metamaterials. Finally, we identified a limit where these differences vanish such that quantum and classical adiabatic and non-adiabatic dynamics become equivalent up to a difference in the dynamic phase. 

\begin{acknowledgments}
Special thanks go to Prof. Gian Michele Graf and Dr. Giovanni Bordiga for helpful discussions. This work was supported by an ETH Zurich Doc.Mobility Fellowship. Furthermore, C. B. and A. F. acknowledge funding from ETH Zurich. M.G. acknowledges support by ERC grant (Project No. 101040117). Views and opinions expressed are however those of the author(s) only and do not necessarily reflect those of the European Union or the European Research Council Executive Agency. Neither the European Union nor the granting authority can be held responsible for them.
\end{acknowledgments}

\bibliography{apssamp}

\appendix

\section{Derivation of equation \eqref{eq: classical mode coupling} and \eqref{eq: classical time evolution operator}} \label{ap: A}
We first derive a few well known auxiliary results that we use later. By differentiating the augmented eigenvalue problem $\tilde{H}(t)\ket{\sigma,m(t)} = \sigma \omega_m(t)\ket{\sigma,m(t)}$ with respect to time and contracting with $\bra{n(t),\tau}$ the following results are obtained:
\begin{equation} \label{ap. eq: auxillary 1}
    \braket{m,\tau|\dot{\tilde{H}}|\sigma,m} = \braket{m|\dot{\sqrt{H}}|,m} \delta_{\sigma \tau} = \dot{\omega}_m \delta_{\sigma \tau},
\end{equation}
secondly,
\begin{equation}\label{ap. eq: auxillary 2}
    \braket{n,\tau|\dot{\tilde{H}}|\sigma,m} = \braket{n|\dot{\sqrt{H}}|m} \delta_{\sigma \tau}= (\omega_m-\omega_n) \braket{n|\dot{m}}\delta_{\sigma \tau},
\end{equation}
and thirdly, by using \eqref{ap. eq: auxillary 1} and \eqref{ap. eq: auxillary 2}
\begin{equation}\label{ap. eq: auxillary 3}
    \braket{n,\tau|\dot{\mathcal{T}}\mathcal{T}^{-1}|\sigma,m} = \begin{cases}
 \frac{1}{2} \frac{\dot{\omega}_n}{ \omega_n}, & \text{if } n = m\\
 \frac{1}{2} \frac{(\omega_m-\omega_n)}{\omega_m}\braket{n|\dot{m}}, & \text{if } n \neq m.
\end{cases}
\end{equation}
Substituting the expansion of equation \eqref{eq: expansion of classical state} in the main text into the classical Schrödinger equation \eqref{eq: classical time depndent schrödinger} and contracting with $\bra{n(t),\tau}$ we obtain
\begin{multline*}
   i \bra{n(t),\tau}\partial_t\Bigg[\sum_{m, \sigma} c_{\sigma,m}(t) \sqrt{ \omega_m(0)\omega_m(t) } e^{-i \sigma \Omega_m(t)}\ket{\sigma, m(t)}\Bigg] \\= \sum_{m, \sigma} c_{\sigma,m}(t) \sqrt{ \omega_m(0)\omega_m(t) } \\
   \times e^{-i \sigma \Omega_m(t)} \bra{n(t),\tau}\Big[\tilde{H}(t)+ i\dot{\mathcal{T}}(t) \mathcal{T}^{-1}(t)\Big]\ket{\sigma, m(t)}.
\end{multline*}
Using the orthonormality $\braket{n(t),\tau|\sigma,m(t)} = \delta_{\tau \sigma} \delta_{nm}$ and the chain rule, the LHS can be expanded as
\begin{multline*}
    \text{LHS} = \\ \Bigg[i\sqrt{\omega_n(0)\omega_n(t)}\dot{c}_{\tau, n} + i \frac{1}{2} \frac{ \sqrt{\omega_n(0)}\dot{\omega}_n}{ \sqrt{\omega_n(t)}}c_{\tau, n} + \tau \sqrt{\omega_n(0)}\omega_n(t)c_{\tau, n} \Bigg] \\ \times e^{-i\tau \Omega_n(t)} \\
    + i \sum_{m\neq n} \sqrt{\omega_m(0)\omega_m(t)}\braket{n|\dot{m}}e^{-i\tau \Omega_m(t)}c_{\tau, m},
\end{multline*}
and the RHS is obtained using the above auxiliary results,
\begin{multline*}
 \text{RHS} = \tau \sqrt{\omega_n(0)} \omega_n(t) e^{-i\tau \Omega_n(t)}c_{\tau, n} \\
 + i \frac{1}{2}\sum_{m \neq n} \sqrt{\omega_m(0)\omega_m(t)}\frac{(\omega_m-\omega_n)}{\omega_m}\braket{n|\dot{m}}e^{-i\tau \Omega_m(t)}c_{\tau, m} \\
 + i \frac{1}{2}\sum_{m \neq n}  \sqrt{\omega_m(0)\omega_m(t)}\frac{(\omega_m-\omega_n)}{\omega_m}\braket{n|\dot{m}}e^{-i(-\tau) \Omega_m(t)}c_{-\tau, m} \\
 + i \sqrt{\omega_n(0)\omega_n(t)}\frac{1}{2} \frac{\dot{\omega}_n}{ \omega_n}e^{-i\tau \Omega_n(t)}c_{\tau, n} \\
 + i \sqrt{\omega_n(0)\omega_n(t)}\frac{1}{2} \frac{\dot{\omega}_n}{ \omega_n}e^{i\tau \Omega_n(t)}c_{-\tau, n}.
\end{multline*}
Multiplying both the LHS and the RHS with $-ie^{i\tau \Omega_n(t)}/\sqrt{\omega_n(0)\omega_n(t)}$ one finds that terms with $c_{\tau,n}$ vanish. By further collecting terms in $\dot{c}_{\tau,n},c_{-\tau,n},c_{\tau,m}$ and $c_{-\tau,m}$ we arrive at 
\begin{multline*}\label{ap eq: classical mode coupling, differential}
\dot{c}_{\tau,n} = \\
- \sum_{m \neq n} \braket{n|\dot{m}} \frac{1}{2} \sqrt{\frac{\omega_m(0)}{\omega_n(0)}} \Bigg(\sqrt{\frac{\omega_n(t)}{\omega_m(t)}} + \sqrt{\frac{\omega_m(t)}{\omega_n(t)}}\Bigg) \\ \times e^{i \tau \Omega^-_{nm}(t)} c_{\tau,m}(t)\\
- \sum_{m \neq n}   \braket{n|\dot{m}} \frac{1}{2} \sqrt{\frac{\omega_m(0)}{\omega_n(0)}} \Bigg(\sqrt{\frac{\omega_n(t)}{\omega_m(t)}} - \sqrt{\frac{\omega_m(t)}{\omega_n(t)}}\Bigg) \\
 \times e^{-i (-\tau) \Omega^+_{nm}(t)}c_{-\tau,m}(t)\\
+  \frac{1}{2} \frac{\dot{\omega}_n(t)}{\omega_n(t)}e^{i\tau 2 \Omega_n(t)}c_{-\tau,n}(t),
\end{multline*}
which can be integrated to yield the integral equation \eqref{eq: classical mode coupling}. We now move on the prove the multiband relations \eqref{eq: time evolution of classical subspace coefficients} and \eqref{eq: classical time evolution operator} in the main text. Under the assumptions stated in equations \eqref{eq: adiabatic condition multiband 1}-\eqref{eq: adiabatic condition multiband 3} the above ODE for the subspace coefficients, $[\dot{\mathbf{c}}^{\tau}_{\mathcal{S}}(t)]_s, s \in \mathcal{S}$,  reduces to 
\begin{multline*}
[\dot{\mathbf{c}}^{\tau}_{\mathcal{S}}(t)]_s = \\ - \sum_{s \neq s'}  \braket{s|\dot{s'}} \frac{1}{2} \sqrt{\frac{\omega_s'(0)}{\omega_{s}(0)}} \Bigg(\sqrt{\frac{\omega_s(t)}{\omega_{s'}(t)}} + \sqrt{\frac{\omega_{s'}(t)}{\omega_s(t)}}\Bigg)
\\
\times e^{i \tau \Omega^-_{ss'}(t)} [\mathbf{c}^{\tau}_{\mathcal{S}}(t)]_{s'},
\end{multline*}
which can be written in matrix-vector notation 
\begin{equation}
        \dot{\mathbf{c}}^{\tau}_{\mathcal{S}}(t) =  \mathcal{C}(t)\Big[A(t)+A^{\Delta}(t)\Big]\mathcal{C}^{-1}(t)\mathbf{c}^{\tau}_{\mathcal{S}}(t),
    \end{equation}
where $\mathcal{C}_{ss'}(t) = \exp(i\Omega_s(t))\delta_{ss'}/\sqrt{\omega_s(0)\omega_s(t)}$, $A_{ss'} = \braket{s|\dot{s}'}$ and $A^\Delta_{ss'}(t) = \braket{s|\dot{s}'}(\omega_s-\omega_{s'})/(2\omega_{s'})$. This directly implies equations \eqref{eq: time evolution of classical subspace coefficients} and \eqref{eq: classical time evolution operator} in the main text by the use of path ordered matrix exponentials \cite{wilczek1984appearance, lam1998decomposition}.

\section{Derivation of equation \eqref{eq: coupling equations for b}} \label{ap: B}
In this appendix we derive equation \eqref{eq: coupling equations for b} in the main text. To that end we use, as in the main text, the following non-adiabatic Ansatz
\begin{multline}
\label{ap eq: expansion of classical state}
\ket{\Psi^c(t)} = \sum_{m, \sigma} b_{\sigma,m}(t)e^{-i \sigma \Omega_m(t)} f_{\sigma,m}(t) \ket{\sigma, M(t)},
\end{multline}
where, as in the main text, $\Omega_m(t) := \int_0^t dt'\omega_m(t')$.
We plug the Ansatz into the equations of motion (equation \eqref{eq: rendered first order} in the main text)
\begin{equation} \label{ap2 eq: rendered first order}
\ket{\dot{\Psi}^c} = G\ket{\Psi^c},
\end{equation}
and contract with $\bra{\hat{N},\tau}1/f^*_{\tau,n}$, where from now on we omit to show the time dependence explicitly. The left hand side is given by 
\begin{multline}
    \braket{\hat{N},\tau|\frac{1}{f^*_{\tau,n}}|\dot{\Psi}^c} \\
    = \dot{b}_{\tau,n} + b_{\tau,n} \frac{\dot{f}_{\tau,n}}{f^*_{\tau,n}} - b_{\tau,n} i \tau \omega_n  +\\
    \sum_{\tau,m}b_{\sigma,m} \frac{f_{\sigma,m}}{f^*_{\tau,n}}\braket{\hat{N},\tau|\sigma, \dot{M}} e^{i(\tau\Omega_n-\sigma\Omega_m)},
\end{multline}
where we used the biorthogonality relation \eqref{ap2 eq: biorthogonality relation} in the main text.
Using the fact that$f_{\sigma,m}(t)\ket{\tau, M(t)}$ are right eigenvectors of $G$, the right hand side can be computed as
\begin{equation}
    \braket{\hat{N},\tau|\frac{1}{f^*_{\tau,n}}G|\Psi^c} = -b_{\tau,n} i \tau \omega_n,
\end{equation}
which is equal to the third term of the LHS. Subtracting it from both sides of the equation and rearranging terms results in 
\begin{multline}\label{ap2 eq: intermediate 0}
     \dot{b}_{\tau,n} = -b_{\tau,n} \frac{\dot{f}_{\tau,n}}{f^*_{\tau,n}} \\
     -\sum_{\tau,m}b_{\sigma,m} \frac{f_{\sigma,m}}{f^*_{\tau,n}}\braket{\hat{N},\tau|\sigma, \dot{M}} e^{i(\tau\Omega_n-\sigma\Omega_m)}.
\end{multline}
Using the orthogonality relation $\braket{n|m} = \delta_{nm}$, we calculate
\begin{multline}
    \braket{\hat{N},\tau|\sigma, \dot{M}} 
    = \frac{1}{2}\Big[\braket{n|\dot{m}}(1+\sigma \tau \frac{\omega_m}{\omega_n})+ \sigma \tau \frac{\dot{\omega}_m}{\omega_n}\delta_{nm} 
    \Big] \\
    =\frac{1}{2} \begin{cases}
        \frac{\dot{\omega}_n}{\omega_n} & m=n, \sigma = \tau \\
        -\frac{\dot{\omega}_n}{\omega_n} & m= n, \sigma \neq \tau \\
        \braket{n|\dot{m}}(1+ \frac{\omega_m}{\omega_n}) & m \neq n, \sigma = \tau \\
        \braket{n|\dot{m}}(1- \frac{\omega_m}{\omega_n}) & m \neq n, \sigma \neq \tau
    \end{cases}.
\end{multline}
Substituting this back into equation \eqref{ap2 eq: intermediate 0} and multiplying by $\sqrt{\omega_n}f_{\tau,n}$ gives
\begin{multline}\label{ap2 eq: coupling equations for b}
    \dot{b}_{\tau,n} f_{\tau,n}\sqrt{\omega_n} = -b_{\tau,n}\dot{f}_{\tau,n}\sqrt{\omega_n} 
    -b_{\tau,n}f_{\tau,n}\frac{1}{2}\frac{\dot{\omega}_n}{\sqrt{\omega_n}} \\
    - \sum_m b_{\tau,m}f_{\tau,m}\braket{n|\dot{m}}\frac{1}{2}\sqrt{\omega_n}\Big(1+ \frac{\omega_m}{\omega_n}\Big) e^{i\tau\Omega^-_{nm}}\\
    - \sum_m b_{-\tau,m}f_{-\tau,m}\braket{n|\dot{m}}\frac{1}{2}\sqrt{\omega_n}\Big(1- \frac{\omega_m}{\omega_n}\Big) e^{i\tau\Omega^+_{nm}}\\
    +b_{-\tau,n}f_{-\tau,n}\frac{1}{2}\frac{\dot{\omega}_n}{\sqrt{\omega_n}}e^{i2\tau\Omega_n},
\end{multline}
which is equation \eqref{eq: coupling equations for b} in the main text. The arguments in the exponentials are defined as in the main text, $\Omega^-_{nm}:=\Omega_n-\Omega_m$ and $\Omega^+_{nm}:=\Omega_n+\Omega_m$.

\section{Single degree of freedom time dependent harmonic oscillator} \label{ap: C}

In this appendix we will apply our study to the single degree of freedom (DOF) harmonic oscillator with time-dependent natural frequency, which is a particular case of a classical system with a single band. Such simple model can be treated using alternative approaches \cite{fiore2022time} and hence provides a comparison reference for the results introduced here.

In \cite{fiore2022time} it is demonstrated that a uniformly converging sequence in $|\dot{\omega}/\omega^2|$ can be obtained for the evolution of the displacement and velocity of the time-dependent single DOF harmonic oscillator. The $0$-th order approximation term of this sequence is given by 
\begin{align}\label{ap3 eq: adiab limit single DOF harmonic oscillator}
        q(t) = \sqrt{\frac{2\mathcal{I}(0)}{\omega(t)}} \sin \Big( \phi(t)\Big), \\ p(t) = \sqrt{2\mathcal{I}(0)\omega(t)} \cos \Big( \phi(t)\Big),
\end{align}
where $q$ and $p$ are the canonical coordinates associated with the Hamiltonian
\begin{equation}
    \mathcal{H}(q,p;t) := \frac{1}{2}\Big(p^2 + \omega^2(t)q^2\Big).
\end{equation}
The phase reads
\begin{equation}
    \phi(t) = \phi(0)+\int_0^t dt' \omega(t')
\end{equation}
and $\mathcal{I}(t) = \mathcal{H}(t)/\omega(t)$. The $0$-th order approximation \eqref{ap3 eq: adiab limit single DOF harmonic oscillator} is valid in the limit where $|\dot{\omega}/\omega^2| \ll 1$. 

Firstly, note that this is the same condition that we obtained in the main text (equation \eqref{eq: classical adiabatic condition 2}, up to a factor $1/4$ which we kept from the derivation for clarity). And secondly, it is easy to see that the depicted evolution \eqref{ap3 eq: adiab limit single DOF harmonic oscillator} is exactly equivalent to equation \eqref{eq: classical adiabatic evolution single band} in the main text for the chosen initial condition of $q(0)=1$ and $p(0)=0$, implying $\mathcal{I}(0)= \omega(0)$ and $\phi(0) = \pi/2$. 

Let us treat the case of general initial conditions given by $q(0)= A \in \mathbb{R}$ and $p(0)=B\in \mathbb{R}$. Note that in the adiabatic limit (or assuming that $\omega(t) = \omega(0)$ for $t\leq0$ and adiabatic changes for $t>0$) this implies that 
\begin{equation}
    \phi(0) = \arctan \Big(\frac{\omega(0)A}{B}\Big).
\end{equation}
This can be derived using the definition $\mathcal{I}(0) = \mathcal{H}(0)/\omega(0)$, $\mathcal{H}(0) = 1/2 (\omega^2(0)A^2 + B^2)$, inserting them into \eqref{ap3 eq: adiab limit single DOF harmonic oscillator} at $t=0$ and requiring that $q(0) = A$ and $p(0) = B$. 

We now demonstrate that we arrive at the same adiabatic dynamics using our approach. The analyis reveals that for general initial conditions both modes are excited but do not interact. 

The equations of motion read
\begin{align}
\ket{\dot{\Psi}^c(t)} = 
    \begin{bmatrix}
        0 & 1 \\ -\omega^2(t) & 0
    \end{bmatrix} 
    \ket{\Psi^c(t)}, \\
    \ket{\Psi^c(0)} = 
    \begin{bmatrix}
        q(0) \\
        p(0)
    \end{bmatrix}
    =\begin{bmatrix}
        A \\
        B
    \end{bmatrix},
\end{align}
and the transformation is given by
\begin{equation}
    \mathcal{T}(t) = \begin{bmatrix}
        \omega(t) & 0 \\
        0 & i
    \end{bmatrix}.
\end{equation}
The non-Hermitian transformed equations of motion for $\ket{\tilde{\Psi}^c(t)} = \mathcal{T}(t)\ket{\Psi^c(t)}$ are 
\begin{align}
    \ket{\dot{\tilde{\Psi}}^c(t)} = 
    \begin{bmatrix}
        \frac{i\dot{\omega}(t)}{\omega(t)}  & \omega(t) \\ \omega(t) & 0
    \end{bmatrix} \ket{\Psi^c(t)}, \\
    \ket{\tilde{\Psi}^c(0)} = \mathcal{T}(0)\begin{bmatrix}
        A \\
        B
    \end{bmatrix} = \begin{bmatrix}
        \omega(0)A \\
        iB
    \end{bmatrix}.
\end{align}

 We have the two frequency subspaces, labelled by $\tau = \pm1$ giving rise to the two modes $\ket{\tau} = 1/\sqrt{2}[1,\tau]^T$ . We therefore make a general, non-adiabatic Ansatz (equation (9) in the main text) of the form
\begin{multline*} \label{ap3 eq: expansion of classical state}
\ket{\tilde{\Psi}^c(t)} = c_{+}(t) \sqrt{ \omega(0)\omega(t) } e^{-i \Omega(t)}\ket{+} \\
+ c_{-}(t) \sqrt{ \omega(0)\omega(t) } e^{i \Omega(t)}\ket{-} ,
\end{multline*}
where $\Omega(t) = \int_0^t dt' \omega(t')$. The initial conditions for the coefficients are given by
\begin{equation}
    c_{\tau}(0) = \braket{\tau|\tilde{\Psi}^c(0)}/\omega(0) = \frac{1}{\sqrt{2}}\Big(A + i\tau B/\omega(0)\Big).
\end{equation}
In the adiabatic limit we have that $c_\tau(t) \approx c_\tau(0)$. The error is given by
\begin{equation}
    c_{\tau}(t) - c_{\tau}(0) = \int_0^t dt c_{-\tau}(t') \frac{1}{2} \frac{\dot{\omega}(t')}{\omega(t')}e^{i\tau 2 \Omega(t')},
\end{equation}
(corresponding to equation \eqref{eq: classical mode coupling} in the main text). From this equation we find that the conditions for the error to be small is given by
\begin{equation}
\max_{t \in [0,T_f]} \Bigg|\frac{1}{4} \frac{\dot{\omega}_n(t)}{\omega^2_n(t)}\Bigg| \ll 1,
\end{equation}
which is the condition \eqref{eq: classical adiabatic condition 2} in the main text and is equivalent to the condition derived in \cite{fiore2022time}
.
Note that this can be viewed as a gap condition for the same mode of the two frequency subspace with a gap of  $|\omega_n-(-\omega_n)| = 2\omega_n(t)$. Or, the alternative intepretation is that the change must be slow compared to how close the mode is from the zero frequency line. Assuming that the condition is satisfied we plug $c_\tau(t) \approx c_\tau(0)$ back into \eqref{ap3 eq: expansion of classical state} apply the inverse transformation, $\mathcal{T}^{-1}(t)$ and extract the displacement:
\begin{multline} \label{ap3 eq: expansion of classical state}
\ket{\psi}^c(t) =   \frac{1}{\sqrt{2}}\Big(A + i B/\omega(0)\Big) \sqrt{ \frac{\omega(0)}{\omega(t)} } e^{-i \Omega(t)}\ket{+} \\
+  \frac{1}{\sqrt{2}}\Big(A - i B/\omega(0)\Big) \sqrt{ \frac{\omega(0)}{\omega(t)} } e^{i \Omega(t)}\ket{-}.
\end{multline}
Taking the real part gives 
\begin{multline}
\psi^c(t) =   \sqrt{ \frac{\omega(0)}{\omega(t)} } \Big(A \cos(\Omega(t)) + B/\omega(0) \sin(\Omega(t)) \Big)\\
= \sqrt{ \frac{\omega(0)}{\omega(t)} } \sqrt{A^2 + (B/\omega(0) )^2}\sin(\phi' + \Omega(t))\\
=  \sqrt{ \frac{\omega^2(0)A^2 + B^2}{\omega(0)\omega(t)} } \sin(\phi' + \Omega(t)),\\
\phi' = \arctan\Big(\frac{A\omega(0)}{B}\Big),
\end{multline} 
which can be identified as equation \eqref{ap3 eq: adiab limit single DOF harmonic oscillator} using that $\mathcal{H}(0) = 1/2 (\omega^2(0)A^2 + B^2)$ and $\mathcal{I}(0) = \mathcal{H}(0)/\omega(0)$. Our analyis illuminates that the non-adiabatic dynamics for a single DOF harmonic oscialator result from the interaction of the two modes $\ket{+}$ and $\ket{-}$. The case of general boundary conditions implies that both modes have a non-zero amplitude but they do not interact. Our analysis further illuminates why the classical time dependent harmonic oscillator can have non-trivial adiabatic dynamics, as there are in fact the two modes at play which can interact and need to be separated to assure adiabaticity. The quantum single DOF case is trivially always adiabatic as there is fundamentally only one mode, and hence no interaction.

\end{document}